\documentclass[a4paper,11pt]{article}
\pdfoutput=1
\usepackage{graphicx}
\usepackage{fullpage}
\usepackage{jheppub} 
\usepackage{bbm}
\usepackage{amsfonts}
\usepackage{slashed}
\usepackage{subfigure}
\usepackage{array}
\usepackage{verbatim}
\usepackage{booktabs}
\usepackage{color, soul}
\usepackage{mathrsfs}
\usepackage{epsfig}
\usepackage{graphicx}
\usepackage{dcolumn}
\usepackage{bm}
\usepackage{amsmath}
\usepackage{amssymb}
\usepackage{multirow}

\setstcolor{red}

\usepackage{makecell}

\def\ep{\epsilon}

\def\bea{\begin{array}}
\def\eea{\end{array}}
\def\beqa{\begin{eqnarray}}
\def\eeqa{\end{eqnarray}}
\def\beqas{\begin{eqnarray*}}
\def\eeqas{\end{eqnarray*}}

\def\bp{\begin{picture}}
\def\ep{\end{picture}}
\def\bc{\begin{center}}
\def\ec{\end{center}}
\def\bfig{\begin{figure}}
\def\efig{\end{figure}}

\def\bit{\begin{itemize}}
\def\eit{\end{itemize}}

\def\f{\frac}

\def\[{\left[}
\def\]{\right]}
\def\({\left(}
\def\){\right)}

\def\..{\left.}
\def\.{\right.}

\def\tm{\times}

\title{\sffamily Explaining the DAMPE data with scalar dark matter and gauged $U(1)_{L_e-L_\mu}$ interaction}

\author[a,b]{Junjie Cao}
\author[c]{,Lei Feng}
\author[a]{,Xiaofei Guo}
\author[a]{,Liangliang Shang}
\author[a,d]{,Fei Wang}
\author[e]{,Peiwen Wu}
\author[c,f]{,Lei Zu}

\affiliation[a]{College of Physics and Materials Science, Henan Normal University, Xinxiang 453007, China}
\affiliation[b]{Center for High Energy Physics, Peking University, Beijing 100871, China}
\affiliation[c]{Key Laboratory of Dark Matter and Space Astronomy, Purple Mountain Observatory, Chinese Academy of Sciences, Nanjing 210008, China}
\affiliation[d]{School of Physics, Zhengzhou University, 450000, ZhengZhou, P.R.China}
\affiliation[e]{School of Physics, KIAS, 85 Hoegiro, Seoul 02455, Republic of  Korea}
\affiliation[f]{School of Astronomy and Space Science, University of Science and Technology of China, Hefei 230026, Anhui, China}

\emailAdd{junjiec@itp.ac.cn}
\emailAdd{fenglei@pmo.ac.cn}
\emailAdd{guoxf@gs.zzu.edu.cn}
\emailAdd{shlwell1988@foxmail.com}
\emailAdd{feiwang@zzu.edu.cn}
\emailAdd{pwwu@kias.re.kr}
\emailAdd{zulei@pmo.ac.cn}

\abstract{
Inspired by the peak structure observed by recent DAMPE experiment in $e^+e^-$ cosmic-ray spectrum, we consider a scalar dark matter (DM) model with gauged $U(1)_{L_e-L_\mu}$ symmetry, which is the most economical anomaly-free theory to potentially explain the peak by DM annihilation in nearby subhalo. We utilize the process $\chi \chi \to Z^\prime Z^\prime \to l \bar{l} l^\prime \bar{l}^\prime$, where $\chi$, $Z^\prime$, $l^{(\prime)}$ denote the scalar DM, the new gauge boson and $l^{(\prime)} =e, \mu$, respectively, to generate the $e^+e^-$ spectrum. By fitting the predicted spectrum to the experimental data, we obtain the favored DM mass range $m_\chi \simeq 3060^{+80}_{-100} \, {\rm GeV}$ and $\Delta m \equiv m_\chi - m_{Z^\prime} \lesssim 14 \, {\rm GeV}$ at $68\%$ Confidence Level (C.L.). Furthermore, we determine the parameter space of the model which can explain the peak and meanwhile satisfy the constraints from DM relic abundance, DM direct detection and the collider bounds. We conclude that the model we consider can account for the peak, although there exists a tension with the constraints from the LEP-II bound on $m_{Z^\prime}$ arising from the cross section measurement of $e^+e^- \to Z^{\prime\ast} \to e^+ e^-$. }

\begin{document}
\maketitle \indent
\newpage

\section{\label{introduction}Introduction}
One of the most important questions in current particle physics and cosmology is to understand the nature of cosmic dark matter (DM).  Although many popular theories can predict viable DM candidates, no DM particle has been discovered by collider or direct detection (DD) experiments so far. So alternative ways, e.g. indirect detection of DM by seeking for its annihilation or decay products, become valuable in understanding the nature of DM.

Recently, the DArk Matter Particle Explorer (DAMPE) experiment released the new measurement of the total cosmic $e^++e^-$ flux between 25 GeV and 4.6 TeV and reported a hint of an excess in the $e^+e^-$ spectrum at around 1.4 TeV \cite{Collaboration2017,DAMPE-1}. Although such an excess may originate from certain new unobserved astrophysical sources, it may also be explained by DM annihilation \cite{DAMPE-4}. Relevant discussion on this subject can be found in \cite{DAMPE-2,DAMPE-3,DAMPE-4,DAMPE-5,DAMPE-6,DAMPE-7,DAMPE-8,DAMPE-9,DAMPE-10,DAMPE-11,DAMPE-12,DAMPE-13,DAMPE-14,DAMPE-15,DAMPE-16,DAMPE-17,DAMPE-18,DAMPE-19}, and one notable conclusion is that if one assumes the $e^+ e^-$ cosmic-ray spectrum to be generated directly from DM annihilation in a nearby clump halo, the best fit values for the DM particle mass, the DM clump mass and the annihilation luminosity are around 1.5 TeV, $10^{6-8} \,M_{\odot}$ and $10^{64-66}\, {\rm GeV^2 cm^{-3}}$, respectively, if the subhalo is about $0.1 \sim 0.3$ kpc away from the earth \cite{DAMPE-4}.

In this work, we attempt to construct a new theory to explain the excess by DM annihilation. For this end, one preliminary requirement on the theory is that the DM annihilation product should be rich in $e^+ e^-$ states. Other requirements include
\begin{itemize}
\label{list-conditions-1}
\item \textbf{I-ID}: current DM annihilation cross section $\langle \sigma v \rangle_0 \gtrsim
1 \times 10^{-26} \,{\rm cm^3/s}$ with $v\sim 10^{-3}\,c$ in the halo,
and meanwhile satisfying the constraints from other DM indirect search experiments.
\item \textbf{II-RD}: DM relic density $\Omega_{\rm DM} = 0.1199\pm 0.0027$ \cite{Planck,WMAP}, which implies that $\langle \sigma v \rangle_{\rm FO} \sim {\cal{O}}(10^{-26}) \, {\rm cm^3/s}$ with $v \sim 0.1 \,c$ in early freeze out.
\item \textbf{III-DD}:  DM direct detection bounds on spin-independent (SI) DM-nucleon scattering rate $\sigma^{SI}_{DM-n}$ from the recent XENON-1T \cite{Aprile:2017iyp}
  and PandaX-II experiments \cite{Cui:2017nnn}.
\item \textbf{IV-Collider}: collider constraints from LHC and LEP-II measurements.
\end{itemize}
The requirements \textbf{I-ID} and \textbf{II-RD} prefer the annihilation process dominated by $s$-wave and at the same time without chiral suppression by light fermion masses \cite{Chang:2013oia,Berlin:2014tja}. Therefore a natural realization to explain the DAMPE excess is the Dirac DM scenario with certain lepton-specific gauge symmetry, where the DM annihilates directly to $e^+e^-$ through $s$-channel mediation of the new gauge boson $Z^\prime$ \cite{DAMPE-2,DAMPE-9,DAMPE-11,DAMPE-13,DAMPE-14}. However, as indicated by recent studies in \cite{DAMPE-2,DAMPE-11,DAMPE-13,DAMPE-14}, even though $Z^\prime$ couples only with leptons, DM-nucleon scattering may still proceed by the $t$-change exchange of the $Z^\prime-\gamma$ transition which is induced by lepton loops. As a result, DM direct detection bounds have tightly limited the parameter space into the resonant annihilation region, $m_{Z^\prime} \simeq 2 m_{{\rm DM}}$, so that it becomes less attractive, especially given the fact that the bound will be improved greatly in near future. On the other hand, the scenario with scalar DM pair annihilation into vector bosons $\chi\chi\rightarrow Z^\prime Z^\prime \to 2 (e^+ e^-)$ can also satisfy the requirements \textbf{I-ID} and \textbf{II-RD}. In our previous work \cite{DAMPE-12} we
studied the scenario with the gauge boson as the mediator between the DM sector and SM sector corresponding to
a gauged family symmetry $U(1)_{B-3(L_e - L_u-L_\tau)}$. We found that the minimal model can not explain the DAMPE results due to the tight constraint from the
LHC search for new gauge boson, but a slightly extended model can do this without conflicting with the constraint \cite{DAMPE-12}.

In principle, the scenario of the scalar DM annihilation with vector portal may be transported to other gauged family symmetries. In this direction,
we are particularly interested in the lepton-specific family symmetry $U(1)_{L_e-L_\mu}$ \cite{He:1991qd,He:1990pn} due to the following considerations.
Firstly, the model is the most economical anomaly-free theory which may explain the peak, so its capability in this aspect
should not be neglected. Moreover, we note that because the gauge boson in the model decays democratically into $e^+ e^-$ and $\mu^+ \mu^-$, instead of
into single $e^+ e^-$ discussed in \cite{DAMPE-14} or equally into three generation lepton pairs discussed in \cite{DAMPE-12},
its prediction on the $e^+ e^-$ spectrum should be different from the previous studies. Secondly, since quarks in the model are uncharged
under the $U(1)_{L_e-L_\mu}$ symmetry, the constraints from LHC experiments and DM direct detection experiments can be greatly relaxed. This
facilitates the model in interpreting the DAMPE excess.
Finally, we note that the model may address the neutrino mass problem \cite{DAMPE-2} and the discrepancy between theoretical prediction and E821
experimental measurement on muon anomalous magnetic momentum \cite{muon:E821}. In this work, we only focus on the features of the model in explaining the DAMPE excess.

The paper is organized as follows. In Section \ref{Section-1}, we show the favored values of $m_{\chi}$ and $m_{Z^\prime}$ by fitting the $e^+e^-$ spectrum
predicted by the DM annihilation $\chi \chi \to Z^\prime Z^\prime \to l \bar{l} l^\prime \bar{l}^\prime$ to the DAMPE results. In Section \ref{Section-2}, we introduce the scalar DM model with $U(1)_{L_e-L_\mu}$ gauge boson as the portal to leptons. In Section \ref{Section-3}, we present the interpretation of the DAMPE excess by  the model while satisfying various experimental constraints. Finally, in Section \ref{Section-Conclusion} we draw our conclusion.

\section{\label{Section-1}Favored masses by the DAMPE excess}

In this section, we determine the mass ranges of $\chi$ and $Z^\prime$ by fitting the $e^+ e^-$ spectrum generated by the process
$\chi \chi \to Z^\prime Z^\prime \to l \bar{l} l^\prime \bar{l}^\prime$ to the DAMPE data. For this end, we first adopt the
parameterization of the cosmic ray background according to the formulae in \cite{DAMPE-7}
and use the package \textbf{LikeDM} \cite{Huang:2016pxg} to calculate the propagation of the $e^+e^-$ flux in the background. The needed
data to determine the parameters for the background are taken as those of the AMS-02 $e^+$ fraction and $e^+e^- $ flux,
and the DAMPE $e^+e^-$ flux \cite{DAMPE-7}.
Then we add the contribution of the local subhalo directly by noting that such a component only affects the energy bin
$\rm \sim 1.5 \,{\rm TeV}$. We adopt a Navarro-Frenk-White profile \cite{Navarro:1996gj} with a truncation at the
tidal radius \cite{Springel:2008cc} for DM density distribution inside the subhalo,
and use the subhalo mass $M_{\rm halo}$ as an input to determine the profile with the method introduced in the appendix of \cite{DAMPE-4}.
The propagation of the nearby $e^+e^-$ can be calculated analytically under the assumption of
spherically symmetric geometry and infinite boundary conditions \cite{Aharonian:1995zz}.
More details of the procedure were introduced in \cite{DAMPE-2,DAMPE-9,DAMPE-12,DAMPE-14}. Finally, we construct a likelihood
function by comparing the predicted spectrum with the AMS-02 data and the DAMPE data, which are distributed from
$0.5 {\rm GeV}$ to $24 {\rm GeV}$ in $36$ energy bins for the former \cite{AMS-02} and from $24 {\rm GeV}$ to $4.57 {\rm TeV}$ in
$40$ energy bins for the latter \cite{Collaboration2017}. Obviously, this
function depends on a series of parameters such as $m_\chi$, $m_{Z^\prime}$, $\langle \sigma v \rangle_0$, $M_{\rm halo}$,
the distance of the subhalo way from the earth $d$ as well as the parameters appearing in the background. Since DAMPE experiment has collected more than thirty
electron/positron events in each energy bin around $1.4 {\rm TeV}$ \cite{Collaboration2017}, we can determine the favored ranges of these parameters by the binned
likelihood fit adopted in this work. For the flux in the $i$th bin, we get its theoretical value by averaging the flux over the width of the energy bin, i.e.
$\Phi_i = \frac{1}{\Delta E} \int \Phi (E) d E$.

Before we proceed, let's illustrate two key features of the spectrum. One is that the height of the spectrum is roughly decided by the product
$\langle \sigma v \rangle_0 \times (M_{\rm halo}/m_\odot)^{0.76}$ \cite{Yuan:2012wj}, which implies that one may fix $\langle \sigma v \rangle_0$ or $M_{\rm halo}$ while varying the other one in performing the fit. Since $M_{\rm halo}$ is also needed to determine the DM profile, in practice
we fix  $M_{\rm halo}$ and vary $\langle \sigma v \rangle_0$ to simplify the calculation.
We note that so far $M_{\rm halo}$ is still unknown and may vary from $10^{6} \,M_{\odot}$ to $10^{8} \,M_{\odot}$ \cite{DAMPE-4},
thus $\langle \sigma v \rangle_0$ can be chosen at the order of $10^{-(24\sim 26)} \,{\rm cm^3 /s}$ in the analysis, which is widely adopted
in many papers. The other feature is that the $e^+ e^-$ spectrum generated by the two-step annihilation $\chi \chi \to Z^\prime Z^\prime \to l \bar{l} l^\prime \bar{l}^\prime$ exhibits box-shaped features,
and the peak-like structure is significant only when the mass splitting $\Delta m \equiv m_\chi - m_{Z^\prime}$ is small.
Since the location and the shape of the spectrum are decided by the parameters $m_{\chi}$ and $\Delta m$, the excess
should impose non-trivial constraints on their values.

\begin{figure}[t]
\begin{center}
\includegraphics[height=6cm,width=8cm]{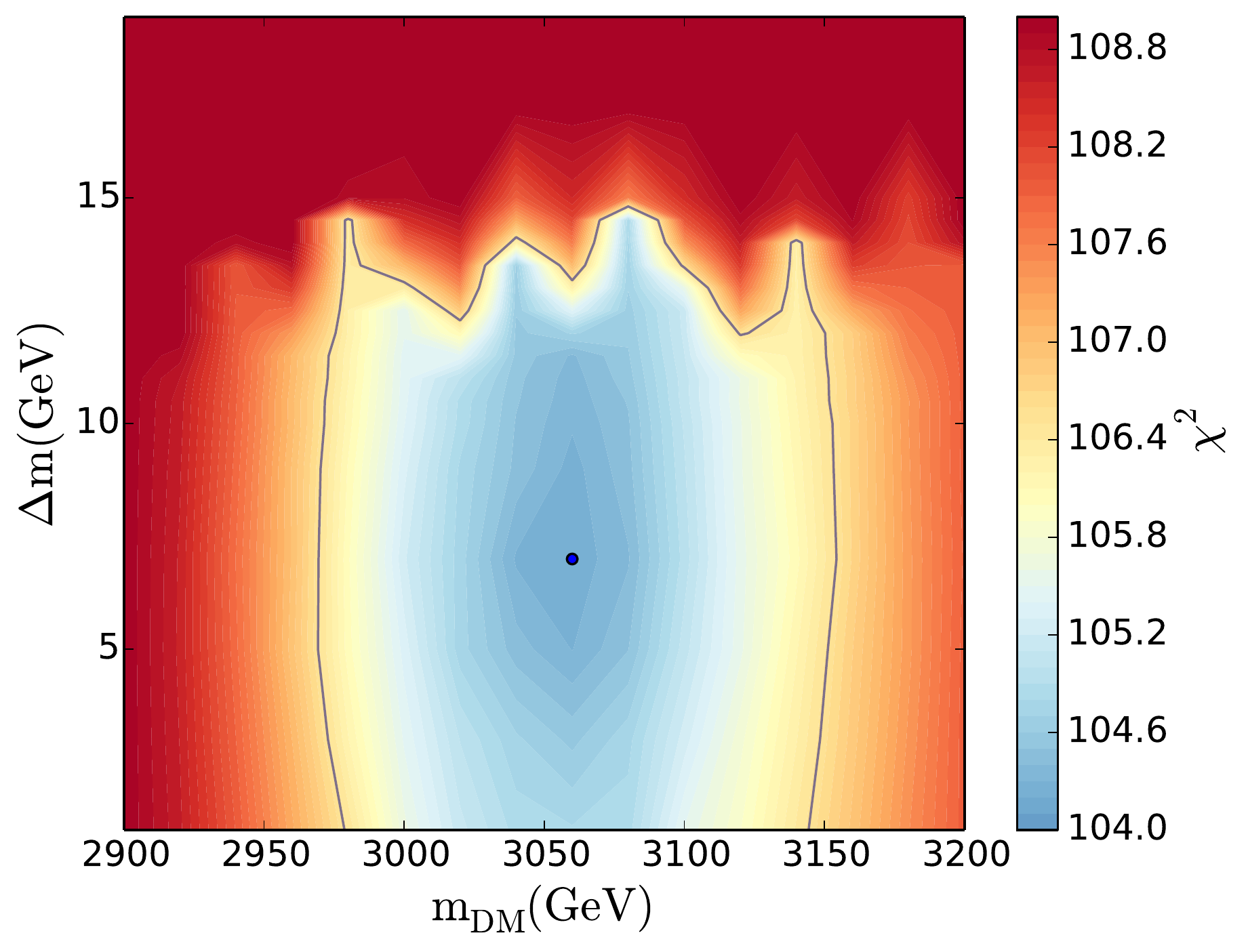}
\includegraphics[height=6cm,width=6cm]{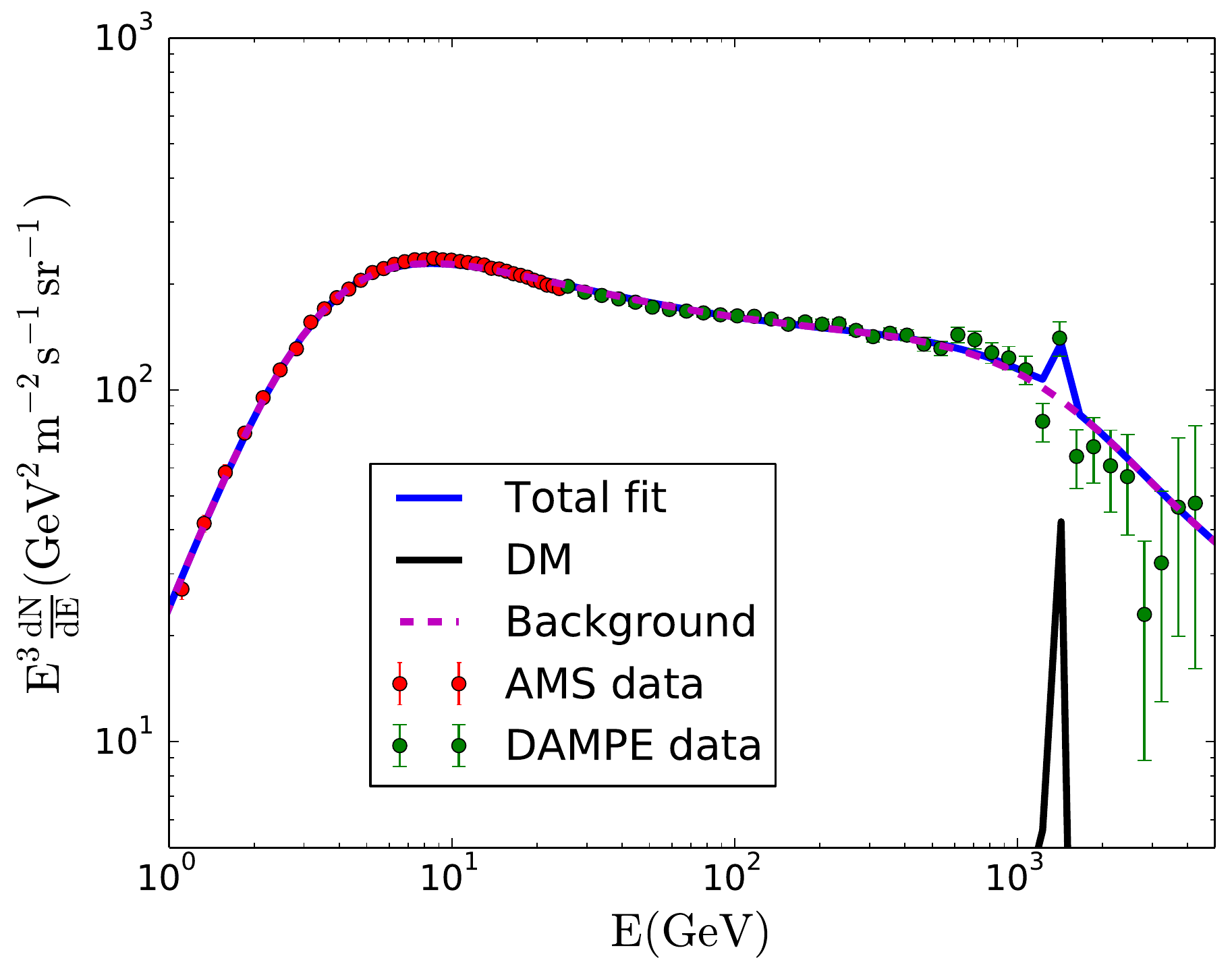}
\caption{Fitting the $e^+e^-$ spectrum generated by the process $\chi \chi \to Z' Z'$ with $Z^\prime \to e^+e^-, \mu^+ \mu^-$ to the AMS-02 and DAMPE data with the background parameterization taken from \cite{DAMPE-7}.
\text{\bf Left} panel: $\chi^2$ map projected on $\Delta m-m_\chi$ plane with $\Delta m \equiv  m_\chi - m_{Z^\prime}$ and the color bar denoting the $\chi^2$ values. The best fit point locates at $(7 \, {\rm GeV}, 3060 \, {\rm GeV})$, and the contour of $\chi^2 = \chi^2_{best} + 2.3$ is also plotted.
\text{\bf Right} panel:  The cosmic $e^+e^-$ spectrum of the best fit point generated by the DM annihilation process in comparison with the AMS-02 and DAMPE data. }
\label{Fit-spectrum}
\end{center}
\end{figure}

In our analysis, we fix $d = 0.1 \,{\rm kpc}$, $M_{\rm halo}= 1.9 \times 10^{7} m_\odot$ and scan the masses $m_{\chi}$ and $\Delta m$.
For each set of $m_{\chi}$ and $\Delta m$, we generate the $e^+e^-$ spectrum arising from both the
background and the signal $\chi \chi \to Z' Z'$ with $Z^\prime \to e^+e^-, \mu^+ \mu^-$, and vary the parameters for the background and $\langle \sigma v \rangle_0$ to get the maximum value
of the constructed likelihood function. Then we obtain the $\chi^2$ of the fit by the formula $\chi^2 = -2 \ln {\cal{L}}_{max}$.
In the \textbf{Left} panel of Fig.\ref{Fit-spectrum}, we show the map of the $\chi^2$ on the $\Delta m - m_{\chi}$ plane
with the color bar denoting the values of the $\chi^2$. We find that the best fit point locates at $(7 \, {\rm GeV}, 3060 \, {\rm GeV})$ with $\langle \sigma v \rangle_0=2.34\times 10^{-26} \,{\rm cm^3/s}$ and its
prediction on the $\chi^2$ is smaller than that obtained from the background-only spectrum by about 5.8. In this panel, we also plot
the constant contour of $\chi^2 = \chi^2_{best} + 2.3$. The region bounded by this contour is interpreted as the best region
of the two-step DM annihilation process to explain the DAMPE excess at $1 \sigma$ level. In the \textbf{Right} panel of Fig.\ref{Fit-spectrum}, we present the
$e^+e^-$ spectrum predicted by the best fit point. In obtaining this panel, we take the parameters for the background from the maximization of the likelihood in
calculating the $\chi^2$. This panel indicates that by choosing appropriate $(\Delta m, m_\chi)$, the process
$\chi \chi \to Z^\prime Z^\prime \to l \bar{l} l^\prime \bar{l}^\prime$ is indeed able to re-produce the DAMPE peak.

\begin{figure}[t]
\begin{center}
\includegraphics[height=6cm,width=6cm]{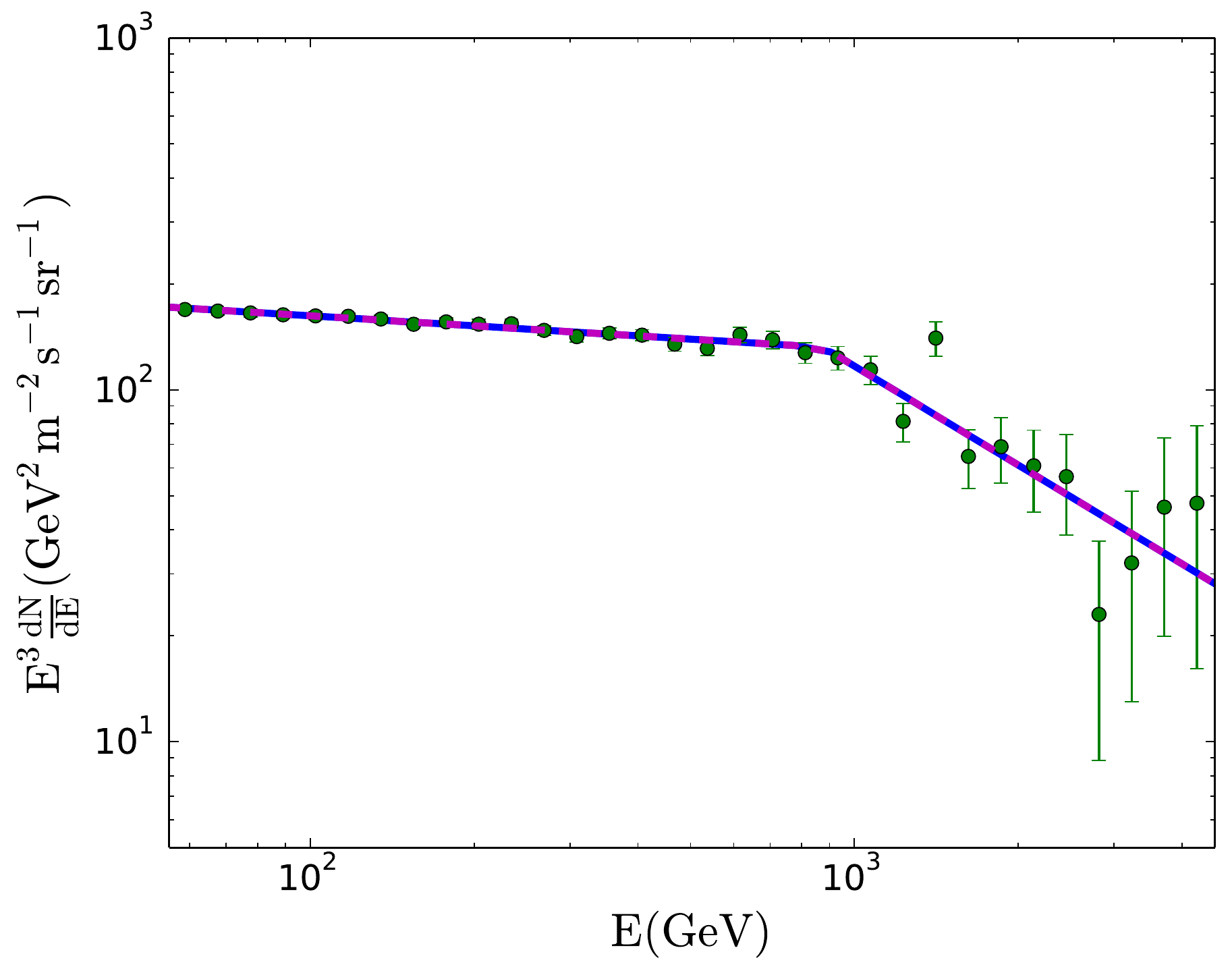}
\includegraphics[height=6cm,width=6cm]{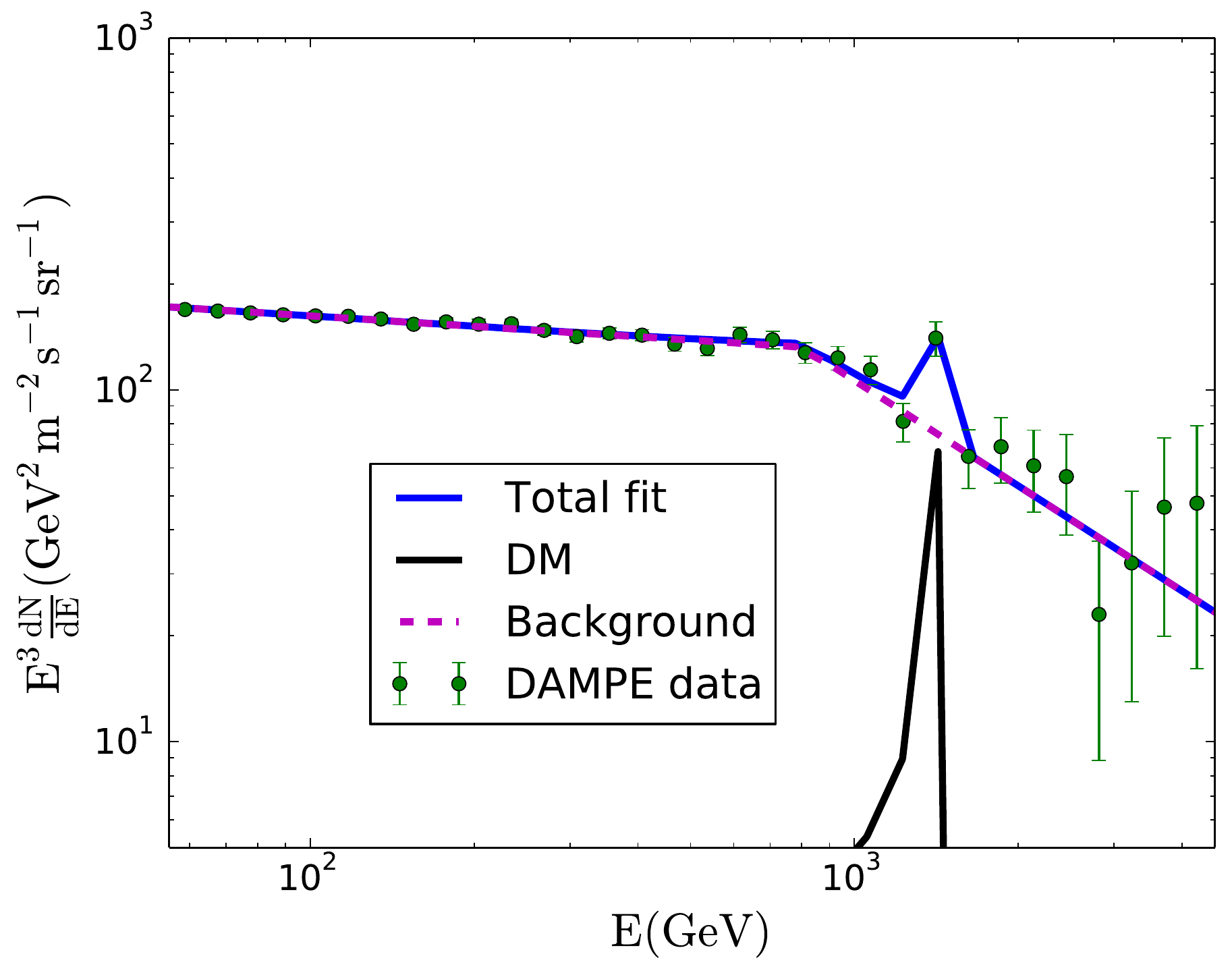}
\caption{Best-fit of the $e^+e^-$ spectrum generated by the process $\chi \chi \to Z' Z'$ with $Z^\prime \to e^+e^-, \mu^+ \mu^-$ to the DAMPE data with the background parameterization taken from \cite{DAMPE-15}. The left panel is based on background-only hypothesis with corresponding $\chi^2 = 25.5$, while the right panel corresponds to background + signal fit with $\chi^2 = 17.0$.  }
\label{Fit-spectrum-2}
\end{center}
\end{figure}

We checked that if we adopt the parameterization of the background by the formulae in \cite{DAMPE-15}, we can roughly reproduce the results of  \cite{DAMPE-15}.
In this case, due to the differences in background parameterization and data set \footnote{In our fit, we varied
six parameters to fit the background and considered total 76 data points with electron/positron energy ranging from $0.5 {\rm GeV}$ to $4.57 {\rm TeV}$
(see the text in the first paragraph of this section). By contrast, the authors in \cite{DAMPE-15} used 4 parameters to reproduce the background and only considered 32 DAMPE points with the energy ranging from $55 {\rm GeV}$ to $4.6 {\rm TeV}$.}, the best point locates at $(8 \, {\rm GeV}, 3061 \, {\rm GeV})$ with
$\langle \sigma v \rangle_0=2.8 \times 10^{-26} \,{\rm cm^3/s}$, and the $\chi^2$ of the best point improves that of
the background-only hypothesis by about 8.5. In Fig.\ref{Fit-spectrum-2}, we plot the best-fit spectrum with the background considered in \cite{DAMPE-15}, and the left and right panels correspond to background-only hypothesis and background + signal hypothesis, respectively. This figure indicates that adopting the parameterizations of the background in \cite{DAMPE-15} can also fit the data well, and does not result in a significant difference in underlying physics.

\section{\label{Section-2}Scalar DM model with gauged $U(1)_{L_e-L_\mu}$ portal}

\begin{table}
\caption{Particle contents and their charges in $G_{SM} \times U(1)_{L_e-L_\mu}$ model.}
\begin{center}
\begin{tabular}{|c|c|c|c|c|c|c|}
\hline
Name 				& Spin	&  Gen. & $SU(3)_C$ & $SU(2)_L$ & $U(1)_Y$  & $U(1)_{L_e-L_\mu}$ \\
\hline \hline
$H$						& 0 		& 1 	& {\bf 1} 					& {\bf 2} 		& -$\frac{1}{2}$ 	&  0 \\
\hline
$Q$						& 1/2 		& 3 	& {\bf 3} 					& {\bf 2} 		&  $\frac{1}{6}$ 	&  0 \\
$d_R^*$ 				& 1/2 		& 3 	& {\bf $\bar{\bf 3}$}	& {\bf 1} 		&  $\frac{1}{3}$ 	& 0 \\
$u_R^*$					& 1/2 		& 3 	& {\bf $\bar{\bf 3}$}	& {\bf 1} 		& -$\frac{2}{3}$	& 0 \\
\hline
$L_1$					& 1/2 		& 1 	& {\bf 1}					& {\bf 2} 		& -$\frac{1}{2}$ 	&  1 \\
$L_2$					& 1/2 		& 1 	& {\bf 1}					& {\bf 2} 		& -$\frac{1}{2}$ 	&  -1 \\
$L_{3}$		         	& 1/2 		& 2 	& {\bf 1} 					& {\bf 2} 		& -$\frac{1}{2}$ 	& 0 \\
$\ell^*_{R,1}$			& 1/2 		& 1 	& {\bf 1} 					& {\bf 1} 		& 1 				& -1 \\
$\ell^*_{R,2}$			& 1/2 		& 1 	& {\bf 1} 					& {\bf 1} 		& 1 				& 1 \\
$\ell^*_{R,3}$	        & 1/2 		& 2 	& {\bf 1}					& {\bf 1} 		& 1 				& 0\\
\hline
$\phi_s$ 				& 0		& 1 	& {\bf 1} 					& {\bf 1} 		& 0 				&  $Y_{s}$ \\
$\phi_\chi$				& 0		& 1 	& {\bf 1} 					& {\bf 1} 		& 0 				&  $Y_{\chi}$\\
\hline
\end{tabular}
\end{center}
\label{table-model}
\end{table}

In this section, we introduce the key features of the scalar DM model which extends the SM gauge group $G_{SM}\equiv SU(3)_C \times SU(2)_L \times U(1)_Y$ by the $U(1)_{L_e-L_\mu}$ gauge symmetry. The particle contents of the model and their charges under the gauged groups are presented in Table \ref{table-model}, where we introduce
a complex scalar $\phi_\chi$ as DM field and a complex scalar $\phi_s$ to be responsible for the broken of the $U(1)_{L_e-L_\mu}$ symmetry as well as for the generation of the $Z^\prime$ mass by its vacuum expectation value (vev) $v_s$ \cite{He:1991qd,He:1990pn}. We impose an odd $Z_2$ parity for $\phi_\chi$ and an even parity for the other fields
to guarantee the stability of the DM candidate. The Lagrangian relevant to our discussion includes
\begin{eqnarray}
\label{eqn-L-model-2}
{\mathcal L} \supset && | D'_{\mu} \phi_\chi |^2 + | D'_{\mu} \phi_s |^2 - V(H,\, \phi_\chi,\, \phi_s) - \frac{1}{4} |F'_{\mu\nu}|^2 \\ \nonumber
&&  + g_{Y^\prime}  Z^\prime_{\mu} ( \overline{e} \gamma^\mu e - \overline{\mu} \gamma^\mu \mu +
\overline{\nu_e} \gamma^\mu \nu_e - \overline{\nu_\mu} \gamma^\mu \nu_\mu ),
\end{eqnarray}
with
\begin{eqnarray}
\label{eqn-L-model-1-V}
V(H,\, \phi_\chi,\, \phi_s) = && m_H^2 |H|^2 + m_{\chi}^2 |\phi_\chi|^2  + m_{s}^2 |\phi_s|^2 \nonumber \\
&& + \lambda_H |H|^4  + \lambda_{\chi} |\phi_\chi|^4 + \lambda_{s} |\phi_s|^4 + \lambda_{\chi H}  |\phi_\chi|^2  |H|^2 \nonumber \\
&& + \lambda_{s H}  |\phi_s|^2  |H|^2 + \lambda_{\chi s}  |\phi_\chi|^2  |\phi_s|^2   + \lambda'_{\chi s}  \Big( (\phi^*_\chi \phi_s)^2 + h.c. \Big). \label{potential}
\end{eqnarray}
In the above expressions, $D'_{\mu} = \partial_\mu - i g_{Y^\prime} Y_\phi Z^\prime_{\mu}$ with $Z^\prime$ denoting the gauge field of the new symmetry, $F'_{\mu\nu}=\partial_\mu Z^\prime_{\nu} - \partial_\nu Z^\prime_{\mu}$, and $m_i$, $\lambda_i$ and $\lambda_{ij}$ with $i,j= H, \chi, s$ are all real free parameters. In Eq.(\ref{eqn-L-model-2})
we neglect the kinematic mixing terms of the two $U(1)$ gauge fields. Note that the $\lambda^\prime_{\chi s}$ term  must vanish if $Y_\chi \neq Y_s $.  Also note that in Eq.(\ref{eqn-L-model-1-V}) we do not consider the special cases of $Y_s=0$ and $2 Y_\chi \pm Y_s = 0$. As a result, cubic operators such as $\phi_s^{(\ast)} \phi_\chi \phi_\chi$ are absent. Moreover, due to the assigned $Z_2$ symmetry, bilinear operators such as $\phi_\chi^\ast \phi_s$ are also absent.

In our discussion, we set $\lambda_{\chi H}=\lambda_{s H}=0$ and choose proper $m_H^2$ and $\lambda_H$ so that the properties of the $H$-dominated scalar are the same as those of the Standard Model (SM) Higgs boson. The minimization conditions of the potential for the fields $\phi_\chi$ and $\phi_s$ are then given by
\begin{eqnarray}
m_\chi^2 + (\lambda_{\chi s} + 2 \lambda^\prime_{\chi s}) v_s^2 + 2 \lambda_\chi v_\chi^2 = 0, \nonumber \\
m_s^2 + (\lambda_{\chi s} + 2 \lambda^\prime_{\chi s}) v_\chi^2 + 2 \lambda_s v_s^2 = 0,
\end{eqnarray}
where $v_\chi$ denotes potential vev of the field $\phi_\chi$. The first condition indicates that for $\lambda_\chi \geq 0$, the field $\phi_\chi$ will not
develop a vev if $m_\chi^2 + (\lambda_{\chi s} + 2 \lambda^\prime_{\chi s}) v_s^2 > 0$.
In this case, $m_s^2 = -2 \lambda_s v_s^2$ and one can replace $m_s^2$ by $v_s$ as a theoretical input parameter. After the symmetry breaking, the terms induced by $\lambda_{\chi s}$ and $\lambda^\prime_{\chi s}$ in the potential can be rewritten as
\begin{eqnarray}
(v_s + \phi_{s, R} )^2 [ (\lambda_{\chi s} - 2 \lambda^\prime_{\chi s} ) \chi^2 + (\lambda_{\chi s} + 2 \lambda^\prime_{\chi s} ) {\chi^{\prime}}^2 ],
\end{eqnarray}
where we define the imaginary and real parts of the field $\phi_\chi$ as $\chi \equiv \phi_{\chi, I}$ and $\chi^\prime \equiv \phi_{\chi, R}$, respectively, and neglect the field $\phi_{s, I}$ since it becomes the longitudinal
component of the gauge particle $Z^\prime$. This expression indicates that $\lambda^\prime_{\chi s}$ can induce a mass splitting between $\chi$ and $\chi^\prime$,
$\Delta m^2_{\chi^\prime \chi} \equiv m_{\chi^\prime}^2 -  m_{\chi}^2 = 8 \lambda_{\chi s}^\prime v_s^2$, and that there is no mixing between $\chi$ and $\chi^\prime$.
Since the parameters $\lambda_\chi$ and $\lambda_{\chi s}$ are unimportant to our discussion, we hereafter treat them as zero .

The gauge charges of the fields $\phi_\chi$ and $\phi_s$ play an important role in DM physics, and some important features about them are listed below.
\bit
\item Because in most cases the DM annihilation $\chi \chi \to Z^\prime Z^\prime$ proceeds mainly via the quartic
$\chi \chi Z^\prime Z^\prime$ coupling, the DM relic density
measured by Planck experiment \cite{Planck,WMAP} and $\langle \sigma v \rangle \propto Y_\chi^4 g_{Y^\prime}^4$ require $Y_\chi g_{Y^\prime} \simeq 1.2$ for $m_\chi \sim m_{Z^\prime} \sim 3 \, {\rm TeV}$ \cite{DAMPE-12}.
This implies that the larger value of $Y_\chi$ one takes, the smaller $g_{Y^\prime}$ should be chosen. In this case, the LEP-II constraints
on $Z^\prime$ mass from $e^+e^- \to Z^{\prime\ast} \to e^+ e^-$ becomes relaxed by the formula $|\frac{g_{Y^\prime} Y_e}{m_{Z^\prime}}| \lesssim 2.02 \times 10^{-4} \, {\rm GeV^{-1}}$ presented in \cite{Schael:2013ita}, where $Y_e = 1$ in our model.

\item If $Y_\chi \neq Y_s $, the $\lambda^\prime_{\chi s}$ term in Eq.(\ref{potential}) must vanish due to gauge invariance.
In this case, $\chi$ and $\chi^\prime$ degenerate in mass and the complex field $\phi_\chi$ acts as the DM candidate. On the other hand, since the
$Z^\prime \phi_\chi^\ast \phi_\chi$ coupling is non-zero, the DM-nucleon
scattering can proceed by the $t$-channel exchange of the $Z^\prime-Z$ transition induced by $e/\mu$ loops
\cite{Kopp:2009et,Huang:2013apa,DEramo:2017zqw}. Consequently, the spin-independent (SI) DM-proton scattering cross section can be approximated by \cite{Berlin:2014tja,DEramo:2017zqw}
\begin{eqnarray}
\sigma_{\phi_\chi -p} \simeq \left \{ Y_\chi^2 \left (\frac{g_{Y^\prime}}{0.6} \right )^4 \left ( \frac{3 \, {\rm TeV}}{m_{Z^\prime}} \right )^2 \right \} \times 0.7 \times 10^{-45} \, {\rm cm^2}.
\end{eqnarray}
This expression indicates that with $Y_\chi g_{Y^\prime} \simeq 1.2$ for $m_{Z^\prime} \simeq 3 \, {\rm TeV}$, $Y_\chi$ must be larger than about 2 in order to survive the tight bound from the DM direct detection experiments
such as XENON-1T \cite{Aprile:2017iyp} and PandaX-II \cite{Cui:2017nnn}. Considering that the bound will be further improved greatly in near future,
we conclude that the case is disfavored if no DM signal is confirmed.

\item If  $Y_\chi = Y_s $, the non-vanishing $\lambda^\prime_{\chi s}$ term can induce a sizable splitting between
$m_{\chi^\prime}$ and $m_{\chi}$, especially $m_{\chi^\prime} > m_{\chi}$ for a positive $\lambda^\prime_{\chi s}$ \cite{DAMPE-12}.
In this case, $\chi$ acts as the DM candidate. More importantly, there is no $Z^\prime\chi\chi$ or $Z^\prime\chi^\prime\chi^\prime$ interaction, but
only $Z^\prime \chi^\prime \chi$ interaction \cite{DAMPE-12}. The remarkable
implication of the mass splitting is that if it is larger than the DM kinetic energy today $\sim 1$ MeV,
the inelastic scattering process $\chi N \to \chi^\prime N$ with $N$ denoting nucleon is kinematically forbidden, while the elastic
scattering process $\chi N \to \chi N$ only starts from 2-loop level since $Z^\prime$ does not couple directly with quarks \cite{DAMPE-12}.
As a result, the DM direct detection experiments will no longer limit the model.

\item In principle, our framework may also account for non-zero neutrino masses and mixing angles
if one further introduces additional scalars \cite{DAMPE-2}. For this purpose, $Y_s$ is naturally chosen to be 2
so that it can couple to the right-handed neutrinos to generate their masses. For such a model, only $Z^\prime$ among the gauge bosons
couples with the right-handed neutrinos, which has important implications at colliders \cite{Das:2017deo,Das:2017flq}.
\eit

Based on the above arguments, we take $Y_\chi = Y_s = 2$ in the following as an example to study the implication of the model in DM physics.

\section{\label{Section-3}Numerical results}

\begin{figure}[t]
\begin{center}
\includegraphics[width=11cm]{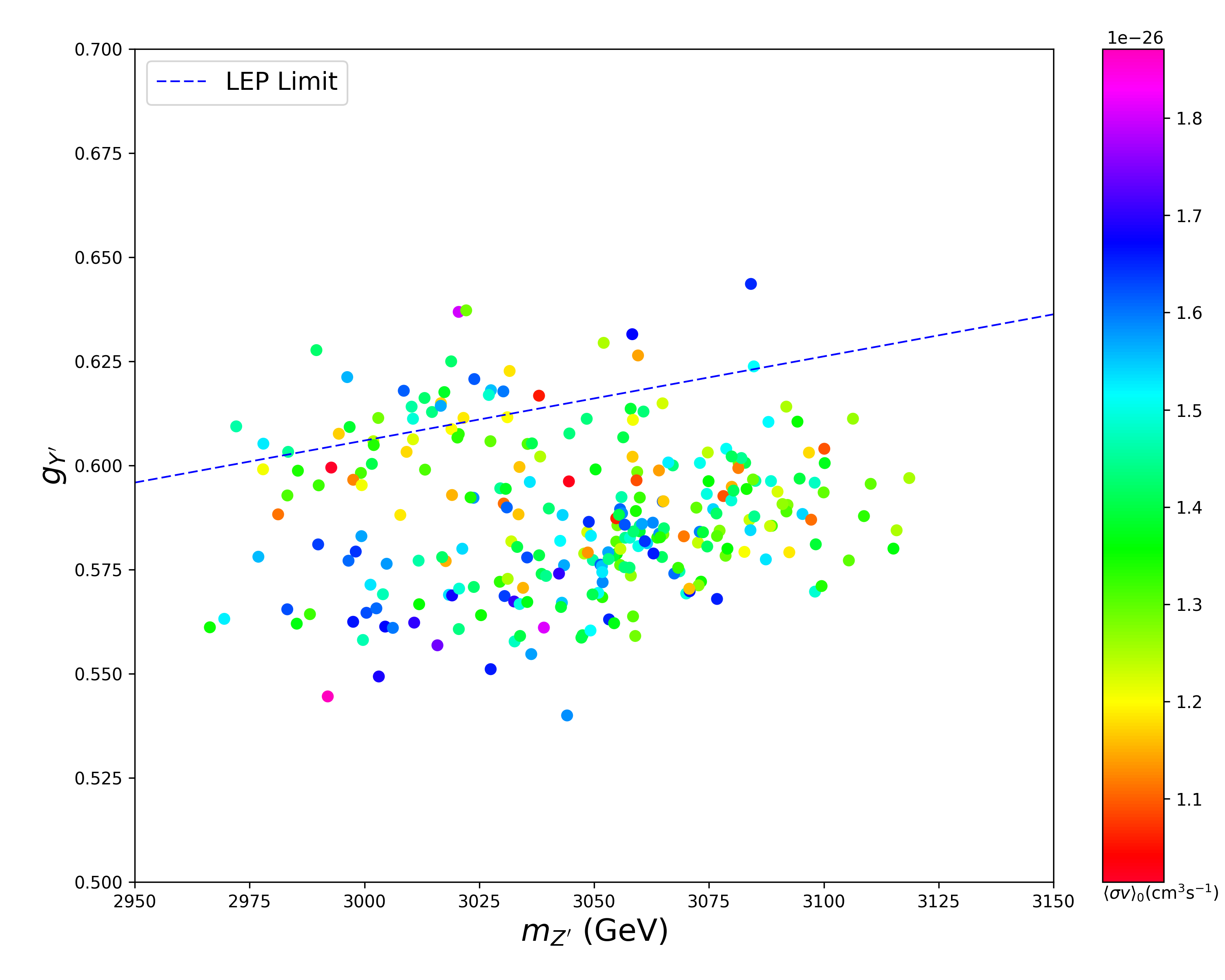}
\caption{Interpretation of the DAMPE excess by the samples satisfying the constraints from DM relic density and direct detection experiments, which are projected on $g_{Y^\prime}-m_{Z^\prime}$ plane. The LEP-II bound on the plane is also plotted with the upper left region of the line excluded. }
\label{fig-model-1-mZp-vs-gZp}
\end{center}
\end{figure}

\begin{table}
\caption{Status of the model $G_{SM} \times U(1)_{L_e-L_\mu}$ confronting the four conditions.}
\begin{center}
\begin{tabular}{|c|c|c|}
\hline
Condition & Result  &  Details \\
\hline \hline
\textbf{I-ID} & $\surd$ & \makecell{$\chi\chi \to Z'Z' $ proceeds mainly by \\
the quartic $\chi \chi Z^\prime Z^\prime$ interaction with $Br(Z^\prime \to e^+e^-) =40\% $
\\ for $m_\chi \sim m_{Z'} \sim 2 \times 1.5$ (TeV) in our model.} \\
\hline
\textbf{II-RD} & $\surd$ & Same as \textbf{I-ID} since the annihilation is a $s$-wave dominated process. \\
\hline
\textbf{III-DD} & $\surd$ & \makecell{ The splitting between $m_{\chi^\prime}$ and $m_\chi$ forbids \\
the inelastic scattering $\chi N \to \chi^\prime N$ by kinematics, \\ while the elastic scattering
 $\chi N \to \chi N$ proceeds at two-loop level. } \\
\hline
\textbf{IV-Collider} & $\surd$ & \makecell{$Y_{\chi}=2$ corresponds to $g_{Y'} \simeq 0.6$ for $m_{Z'}\simeq 3$ TeV, which
is \\ on the edge of being excluded by the LEP constraint on $m_{Z^\prime}$. \\ A larger
$Y_\chi$ can alleviate the tension. Since $Z^\prime$ does not couple \\
with quarks, LHC experiments do not provide any constraints. } \\
\hline
\end{tabular}
\end{center}
\label{table-model-summary}
\end{table}

In the numerical calculations, we use the public package \textbf{SARAH} \cite{Staub:2015kfa} to implement the model, \textbf{SPheno} \cite{Porod:2003um,Porod:2011nf} to obtain the mass spectrum and \textbf{micrOMEGAs} \cite{Belanger:2014vza,Belanger:2014hqa} to calculate the DM relic abundance in which the threshold effects may be important when $m_\chi\sim m_{Z'}$ \cite{PhysRevD.43.3191}. We scan the following parameter space by the package \textbf{Easyscan-HEP} \cite{Easyscan} which is based on Markov Chain Monte Carlo (MCMC) sampling technique \cite{MCMC},
\begin{eqnarray}
0 \leq \lambda_{s},\ g_{Y^\prime}, \,  \lambda^\prime_{\chi s} \leq 1, \  2.9 \,{\rm TeV} \leq m_{\chi} \leq 3.2 \,{\rm TeV}, \
1 \,{\rm TeV} \leq v_s \leq 5 \,{\rm TeV}.
\end{eqnarray}
In the scan, we impose the constraints (see the four conditions listed in Section I)
\begin{eqnarray}
m_{\phi_{s,R}},m_{\chi'} > 3.5 \,{\rm TeV}, \quad \Omega_{\chi} h^2 &\in& 0.1187 \pm 0.01198, \quad
\langle \sigma v \rangle_0 \gtrsim 1\times 10^{-26}\, {\rm cm^3/s}, \label{Constraints}
\end{eqnarray}
where both of the two physical scalars $\phi_{s,R}$ and $\chi^\prime$ are required to
be heavier than $3.5 \,{\rm TeV}$ so that they do not affect the DM annihilation. Moreover, we
require that $m_{\chi}$ and $\Delta m$ are in the region bounded by the constant $\chi^2$
contour in Fig.\ref{Fit-spectrum}.

\begin{figure}[t]
\begin{center}
 \includegraphics[width=7cm]{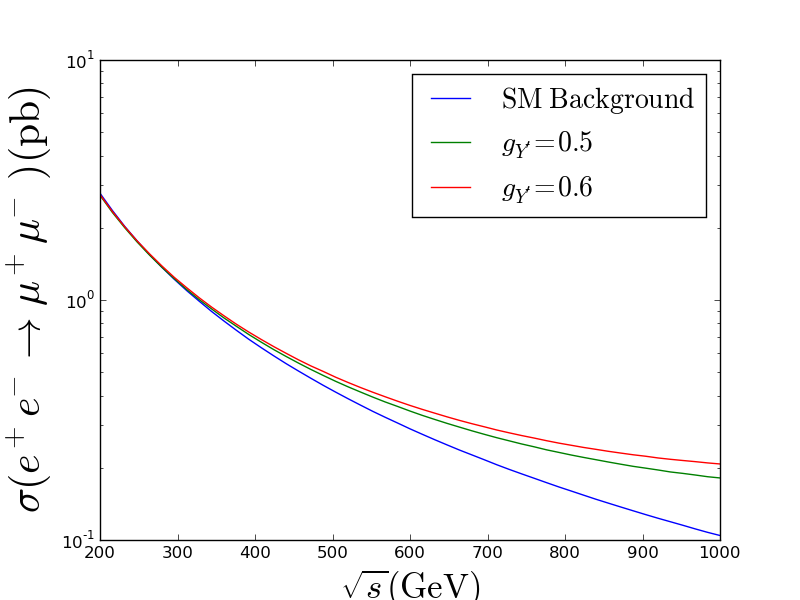}
\includegraphics[width=7cm]{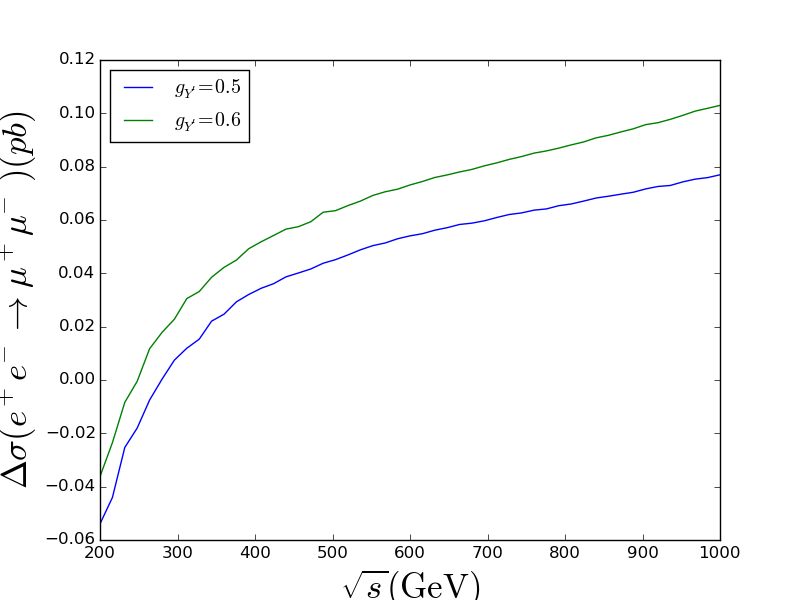}
\caption{Modified production rate $\sigma(e^+e^- \to \mu^+\mu^-)$, which is relevant in searches for $Z'$ at ILC. }
\label{fig-model-1-s-vs-sigma}
\end{center}
\end{figure}

The numerical results are provided in Fig.\ref{fig-model-1-mZp-vs-gZp} on the plane of  $g_{Y^\prime}$ versus $m_{Z^\prime}$, where
the upper left region of the LEP-II bound, i.e. $\frac{g_{Y'}}{m_{Z'}} \gtrsim 2.02 \times 10^{-4} \, {\rm GeV^{-1}}$ \cite{Schael:2013ita}, has been
excluded. This figure indicates that the LEP-II bound has tightly limited the ability of the model to explain the DAMPE excess, nevertheless
there still exist surviving samples that are capable of the explanation. As we discussed in previous section, the tension can be relaxed greatly if one adopts a larger $Y_\chi$. For example, if one chooses $Y_\chi^\prime =3 $, the observed relic density and the interpretation of the DAMPE excess require $g_{Y^\prime} \simeq 0.4$ for $m_{Z^\prime} \simeq 3 \, {\rm TeV}$, in which case the LEP-II experiments has no
constraint on the model.  Moreover, it should be noted that since the SM quarks are singlets under the group $U(1)_{L_e-L_\mu}$, there are no Drell-Yan constraints on $Z^\prime$ from the LHC \cite{Aaboud:2017buh}, which is different from our previous work \cite{DAMPE-12}.
In order to illustrate clearly the status of our explanation
confronting the four conditions listed in Section I, we summarize the details in Table \ref{table-model-summary}.

Before we end the discussion, we have the following comments.
\begin{itemize}
\item As was pointed out in \cite{DAMPE-2, DAMPE-14}, the DM explanation of the DAMPE peak is consistent with other DM indirect detection experimental results such as the H.E.S.S. data on the annihilation $\chi \chi \to V V \to 2 (e^+ e^-) $ \cite{Abdallah:2016ygi,Profumo:2017obk}, the Fermi-LAT data in the direction of the dwarf spheroidal galaxies \cite{Ackermann:2015zua}, the Planck CMB data which is sensitive to energy injection to the CMB from DM annihilations \cite{Slatyer:2015jla,Slatyer:2015kla}, and the IceCube data on DM annihilation into neutrinos \cite{Aartsen:2017ulx}. It also survives the upper bounds from XENON-10 and XENON-100 experiments on the DM scattering off electron \cite{Essig:2017kqs}.
\item The DAMPE explanation imposes non-trivial requirements on $g_{Y^\prime}$ and $m_{Z^\prime}$ which will be further tested at the future International Linear Collider (ILC). For example, it is expected that the sensitivity $\frac{g_{Z'}}{m_{Z'}}\gtrsim 2.2 \times 10^{-5} \, {\rm GeV^{-1}}$ will be achieved
    at the ILC with $\sqrt{s}= 1 \, {\rm TeV}$ and a luminosity of 500 ${\rm fb^{-1}}$  \cite{Freitas:2014jla}. To gain a more complete picture, we also show the modified production cross section $\sigma(e^+e^- \to \mu^+\mu^-) $ in Fig.\ref{fig-model-1-s-vs-sigma} for $m_{Z^\prime}=3$ TeV with two benchmarks $g_{Y^\prime}=0.6, 0.5$. This figure indicates that the gauge boson $Z^\prime$ can indeed make sizable difference in the production rate at the ILC.

\item In our scalar DM model, $Z^\prime$ contributes to the muon anomalous magnetic momentum $g-2$ by \cite{muon:g-2}
  \beqa
  \Delta a_\mu\approx\f{3g_{Y^\prime}^2}{4\pi^2}\f{m_\mu^2}{M_{Z^\prime}^2}.
  \eeqa
  For $g_{Y^\prime} \simeq 0.6$ and $m_{Z^\prime} \simeq 3 \, {\rm TeV}$, this contribution is about $3.35\tm 10^{-11}$ which is too small to account
  for the measured value of muon $g-2$.

\item Throughout this work, we have chosen the scalar DM model with the $U(1)_{L_e-L_\mu}$ symmetry to explain the DAMPE excess.
  Alternatively, one may also choose a similar model but with $U(1)_{L_e-L_\tau}$ gauge symmetry. We find that roughly the same results
  for the $e^+e^-$ spectrum and the favored $(\Delta m, m_\chi)$ region as those in Fig.\ref{Fit-spectrum} can be obtained. Therefore we
  conclude that the excess may also be explained by the $U(1)_{L_e-L_\tau}$ symmetry.
\end{itemize}

\section{\label{Section-Conclusion}Conclusion}

In this work, we discussed the feasibility of scalar DM annihilation to explain the tentative peak observed by the recent DAMPE experiment in the cosmic $e^+e^-$ flux.  Assuming that the two-step process $\chi \chi \to Z' Z' \to \ell\bar{\ell}\ell'\overline{\ell'}$ with $\ell^{(')}=e,\mu$ is responsible for the $e^+ e^-$ spectrum, we first determined by fitting the spectrum to the DAMPE data the favored masses of the DM $\chi$ and the mediator $Z^\prime$, which are presented on $\Delta m-m_\chi$ plane in Fig.\ref{Fit-spectrum}. Then we considered a scalar DM model with gauged $U(1)_{L_e-L_\mu}$ symmetry to realize the process and investigated the features of the model in explaining the excess. Our results indicate that the model can account for the excess without conflicting constraints from DM direct and indirect detection experiments as well as collider experiments. Depending on the assignment of DM charge $Y_\chi$ under the $U(1)_{L_e-L_\mu}$ symmetry, one can alleviate the tension between the DAMPE interpretation and the LEP-II bound on $m_{Z^\prime}$. These observations are presented in Fig.\ref{fig-model-1-mZp-vs-gZp} and also summarized in Table \ref{table-model-summary}.

Before we end this work, we want to emphasize some facts. First, the scalar DM model with the $U(1)_{L_e-L_\mu}$ symmetry is one of the most economical anomaly-free theory which may be used to explain the DAMPE excess. So its features in this aspect should be investigated carefully, especially in light of the fact that another economical anomaly-free $U(1)$ extension of the SM studied in \cite{DAMPE-12} fails in doing this.  As we have shown in this work, the theory can explain the excess without conflicting with any experimental measurements, and a large $Y_\chi$ is favored to relax the tension with the LEP-II constraint. These conclusions are rather new. Second, as we discussed in Section II, the measured $e^+e^-$ flux constrain nontrivially the parameters space of $m_\chi$ and $m_{Z^\prime}$. In getting the $\chi^2$ map on $\Delta m - m_\chi$ plane in Fig.\ref{Fit-spectrum}, we solved the propagation equation of electron/positron in cosmic ray and sought for the maximum value of the likelihood function, which depends on seven variables, for each set of ($m_\chi$,$\Delta m$). So the involved calculation is rather complex and time consuming. We remind that, as far as we know, such a $\chi^2$ map was not obtained in previous
literatures to explain the DAMPE excess.
Finally, we'd like to clarify the relation of this work with our previous work \cite{DAMPE-12}, where we utilized another $U(1)$ extension of the SM to explain the excess. In either work, a scalar DM candidate from a minimal and anomaly-free theory was considered to generate the $e^+e^-$ flux by the two step annihilation process $\chi \chi \to Z^\prime Z^\prime \to l \bar{l} l^\prime \bar{l}^\prime$. As a result, the scalar potentials in the two works and also the relevant experimental constraints are same, and therefore we organize the discussions in a similar way. However, the underlying physics is quite different, which can be seen in the following aspects:
\begin{itemize}
\item the decay products of the mediator $Z^\prime$, or equally speaking the final state in DM annihilation. For the theory considered in \cite{DAMPE-12}, $Z^\prime$ decays democratically into $e^+ e^-$, $\mu^+ \mu^-$ and $\tau^+ \tau^-$ to explain the excess. This was motivated by the original work \cite{DAMPE-4} about the DAMPE experiment,  where the authors pointed out by simulation that pure $e^+e^-$ final state or equally mixed $e^+e^-$, $\mu^+\mu^-$ and $\tau^+\tau^-$ final state is capable of predicting the right shape and height of the spectrum. In this work, however, $Z^\prime$ decays with an equal rate into $e^+ e^-$ and $\mu^+ \mu^-$ states, and  whether the final states can explain the excess is still unknown before our study. We stress that different final states can result in significant difference in the $e^+e^-$ spectrum, and consequently can favor different ranges of physical inputs. For example, as was illustrated in a later  discussion in \cite{DAMPE-19},  the former state prefers a larger mass splitting and a larger annihilation cross section than the latter state.
\item more important, the status of the theory confronting the constraints. Explicitly speaking, the anomaly-free condition for the former theory requires that the quark fields in the SM carry a charge of $\frac{1}{3}$ under the new $U(1)$ symmetry. This charge assignment can induce the tree-level process $p p \to Z^\prime \to l^+ l^-$ at the LHC, and the analysis by ATLAS collaboration on dilepton signal has pushed the lower mass bound of $Z^\prime$ up to about $4$ TeV, which implies that the minimal framework fails to account for the excess \cite{DAMPE-12}. By contrast, the explanation in this work is free of such a problem.
\end{itemize}

\section*{Acknowledgement}

This work is supported by the National Natural Science Foundation of China (NNSFC) under grant No. 11575053,11675147,
 by the Innovation Talent project of Henan Province under grant number 15HASTIT017, by the Youth Innovation Promotion Association CAS (Grant No. 2016288),
 by the Young Core instructor of the Henan education department.


\bibliographystyle{JHEP}
\bibliography{DAMPE-LeLmu}

\end{document}